\documentclass[11pt,twoside]{article}

\usepackage{asp2006}
\usepackage{epsf}
\usepackage{psfig}
\usepackage{lscape}

\begin{document}

\newcommand{\DeltaD}{{$\Delta D$}}
\newcommand{\DeltaV}{{$\Delta V$}}
\newcommand{\Nhost}{{\rm N}}
\newcommand{\Nhalo}{{\rm N}}
\newcommand{\DN}{${\rm D_N}$}
\newcommand{\ewha}{${\rm EW(H\alpha)}$}
\newcommand{\Vin}{${\rm V_{in}}$}
\newcommand{\Vnow}{${\rm V_{now}}$}\
\newcommand{\Vmax}{${\rm V_{\rm max}}$}
\newcommand{\Nseven}{${\rm N_{700}}$}
\newcommand{\kms}{km s$^{-1}$}
\newcommand{\kpc}{h$^{-1}$ kpc}
\newcommand{\Mrh}{M$_{\rm r} + 5 \log{\rm h}$}

\title{Tracking Evolutionary Processes with Large Samples of Galaxy Pairs}   
\author{Elizabeth J. Barton, Christopher Q. Trinh, James S. Bullock, and Shelley A. Wright\altaffilmark{1}}   
\affil{Center for Cosmology, University of California, Irvine}   
\altaffiltext{1}{Present Affiliation: University of California, Berkeley}
\begin{abstract} 
Modern redshift surveys enable the identification of large samples of
galaxies in pairs, taken from many different environments.  Meanwhile,
cosmological simulations allow a detailed understanding of the
statistical properties of the selected pair samples.  Using these
tools in tandem leads to a quantitative understanding of the effects
of galaxy-galaxy interactions and, separately, the effects quenching
processes in the environments of even very small groups.  In the era
of the next generation of large telescopes, detailed studies of
interactions will be enabled to much higher redshifts.
\end{abstract}


\section{Introduction}   

In many ways, the early stages of galaxy-galaxy interactions are the easiest to
identify and study.  When they exist as a resolved pair, large samples
of hundreds or thousands of pre-merger galaxies can be selected
readily and objectively from a redshift survey
\citep[e.g.,][]{barton00, lambas03, nikolic04}.  This technique has a
major advantage: subjective, morphological, and/or kinematic measures
of whether two galaxies are merging are completely unnecessary.
Because the common indicators of interaction depend on merger stage
and surface brightness, and because they can miss retrograde
encounters and minor interactions, samples selected based on merger
signatures are less complete than pair samples.  

Unfortunately, pair samples include significant contamination from
pairs that are physically well separated and seen projected, pairs
that are not physically close but are part of the same larger group,
and pairs that have not yet interacted.  In \citet{barton07}, we use
cosmological $N$-body simulation coupled with a semianalytic
substructure model \citep{zentner03,zentner05} to explore the
contamination inherent in pair samples selected from redshift surveys.
Here, we describe that technique and extend it to study
star formation suppression in small groups.

\section{Close Pairs and Triggered Star Formation}   

\begin{figure}
\plottwo{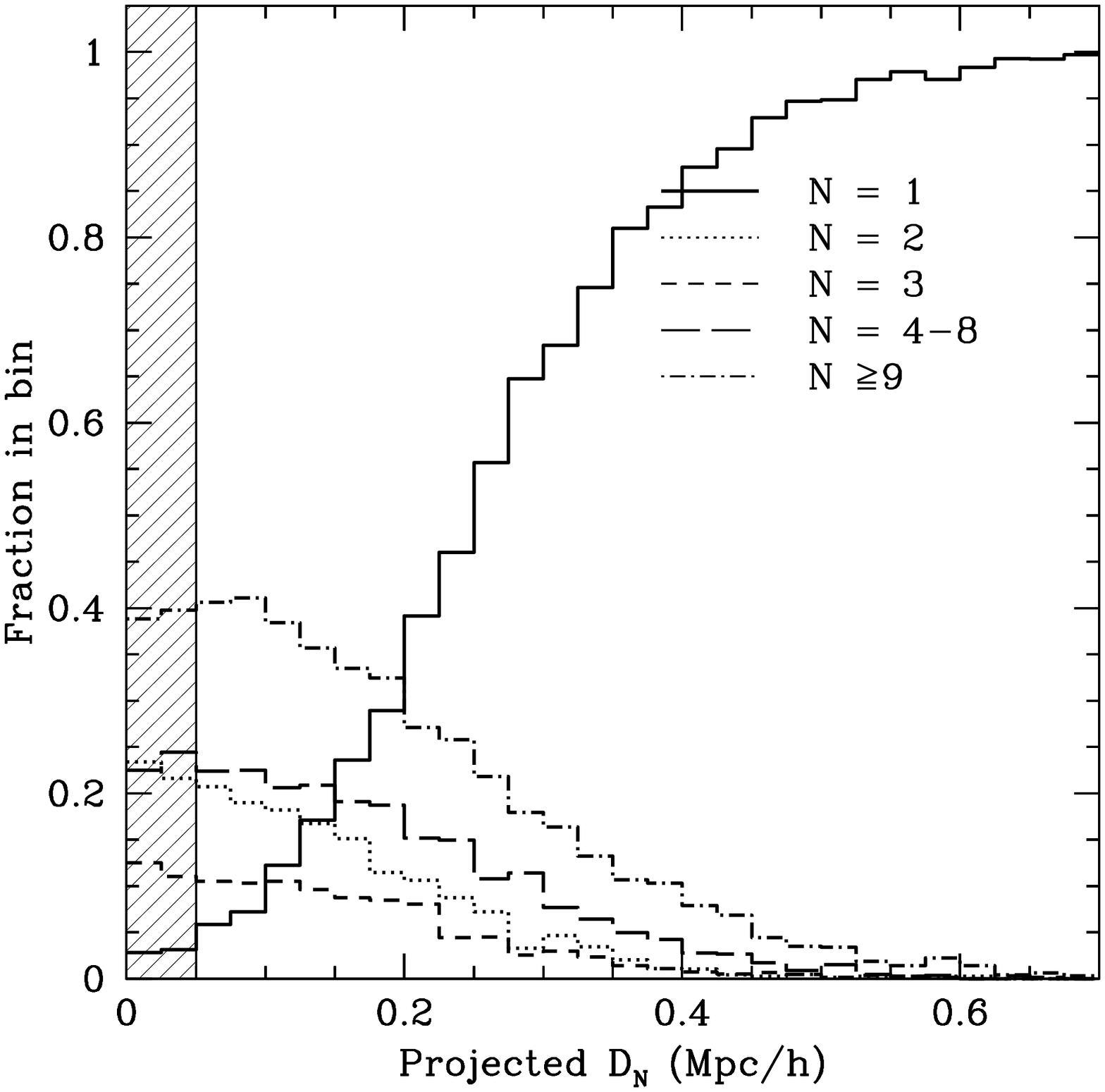}{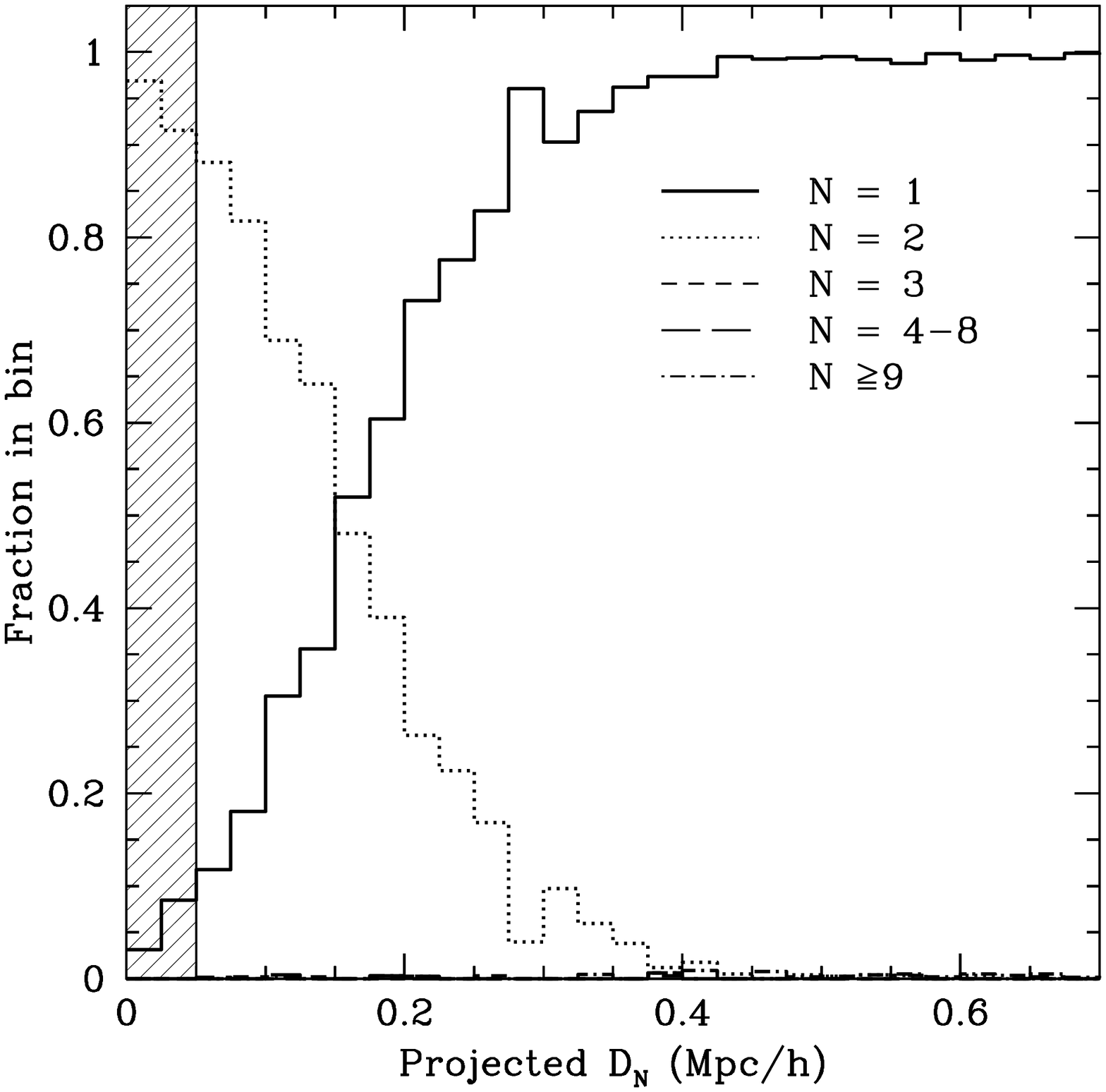}
\caption{Properties of pairs in a cosmological simulation, from
\citet{barton07}.  Using a cutoff halo maximum velocity of 160.5
km s$^{-1}$, we bin the data by the projected distance to the nearest
neighbor, \DN, within 1000 km$^{-1}$ and plot the fraction of galaxies in each bin
with a given of the number of massive subhalos, \Nhalo\ (including the central
galaxy).  We plot the full sample ({\it Left}) and the sample with at
most one neighbor within 700 h$^{-1}$ kpc ({\it Right}).  If no
isolation criteria are applied, apparent pairs are {\it mostly}
galaxies in larger groups typically with $\geq 4$ members.  If the
isolation criterion is applied, close pairs are primarily isolated
\Nhalo\ $=2$ systems.}
\label{fig:sims}
\end{figure}

\citet{barton07} contains a detailed description of our technique for
isolating triggered star formation in galaxy pairs.  The essence of
the approach is to identify galaxy pairs that are unlikely to be part
of larger and more complex systems, to estimate the likely
contamination from isolated galaxies seen in projection, and to
identify an appropriate control sample of isolated galaxies for direct
comparison.  We use the cosmological model to understand galaxies in
pairs.  Specifically, we create an artificial volume-limited redshift
survey from the hybrid $N$-body simulation and substructure model by
matching the abundances of simulated dark matter halos and sub-halos
to the abundances of the galaxies in the redshift survey.  This
technique rests only on the assumption that galaxy luminosity depends
monotonically on halo mass or, in our case, on the surrogate maximum
circular speed of the simulated halo or sub-halo, V$_{\rm max}$.
For sub-halos, we use the circular speed of the dark matter halo at the time it fell into the
parent halo
because it is consistent with the two-point correlation function and
pair counts statistics of existing redshift surveys \citep{conroy06,berrier06}.

Fig.~\ref{fig:sims}a employs this technique, showing the primary
limitation of pair samples selected from redshift surveys.  For a
volume-limited sample that corresponds to $M_{B,j} \leq -19$ in the
2dF Galaxy Redshift Survey \citep{colless01}, we show that most
galaxies in close pairs are actually members of larger groups of 3 or
more galaxies. As a result, many galaxies in statistical pair samples
are actually cluster or group galaxies, which likely have suppressed
star formation for other reasons.  This effect causes a strong and
systematic underestimate of the aggregate amount of star formation in
close galaxy-galaxy pairs.  This effect, and not dust, is likely
responsible for the result that somewhat widely separated pairs
($\sim$60-100 h$^{-1}$ kpc) are actually {\it redder} than the field
\citep[e.g.,][]{alonso06}.

\begin{figure}
\epsscale{0.5}
\plotone{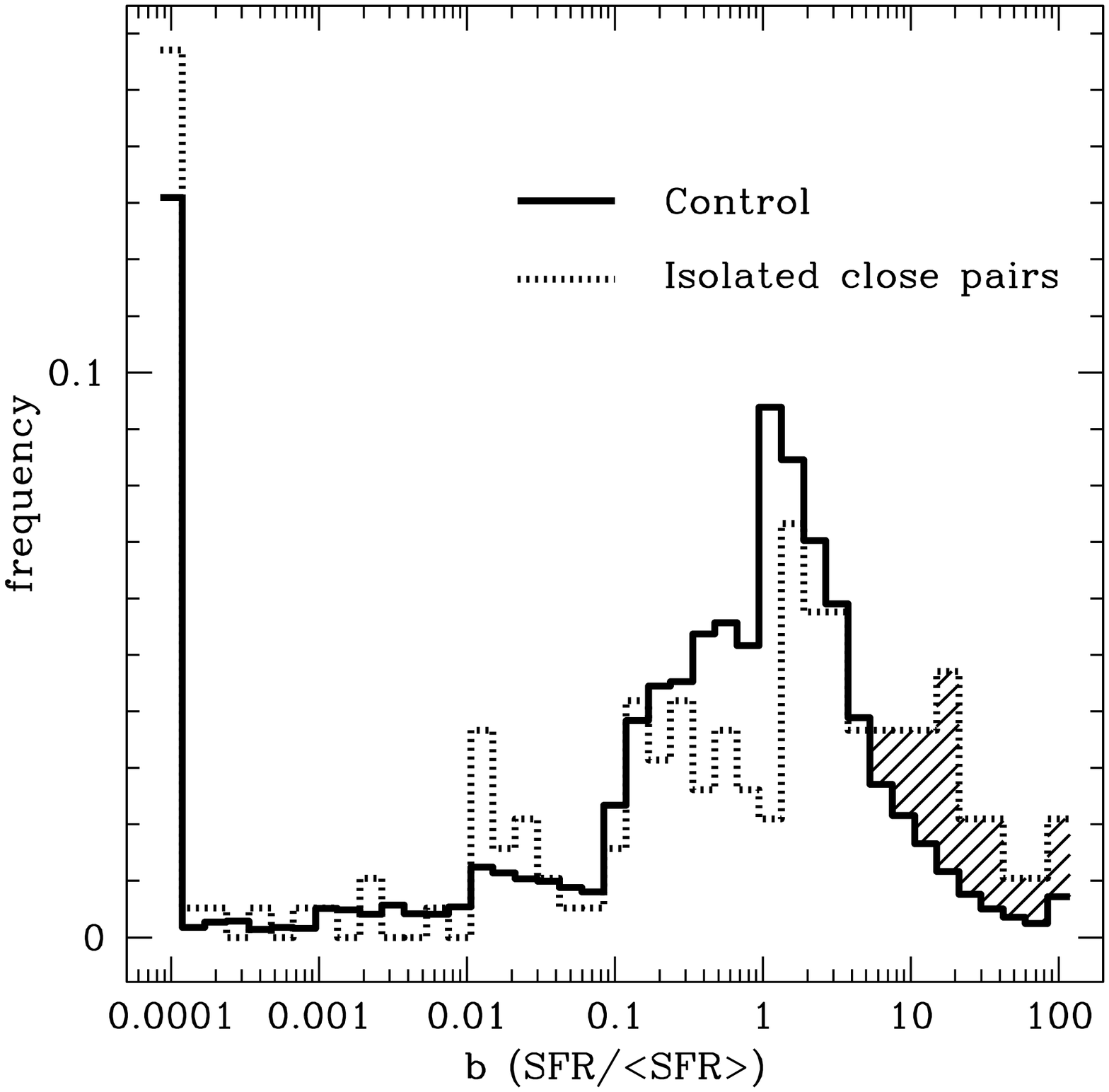}
\caption{A fair comparison of isolated galaxies ({\it Solid line}) and
isolated close pairs ({\it Dashed line}), from \citet{barton07}.  We plot histograms of the 
star formation rate divided by the average past star formation rate, as estimated
from the 2dF spectra \citep[see][]{colless01, madgwick03,barton07}, for the isolated
galaxies and the isolated pairs.  The isolated pairs show a significant (14\%) excess of 
star-forming galaxies with rates boosted by  $b \geq 5$ ({\it Shaded histogram}).}
\label{fig:2dF}
\end{figure}

If we restrict the sample to galaxies with at most one neighbor within
700 h$^{-1}$~kpc on the sky and 1000~km s$^{-1}$ in redshift, however,
as we do in Fig.~\ref{fig:sims}b, it becomes straightforward to
identify isolated galaxy pairs.  For pairs with separations $\leq
50$~h$^{-1}$ kpc, the contamination in the pair sample from isolated
(\Nhalo\ $=1$) galaxies is $\leq 10$\% and the contamination from
systems with $\geq 3$ members is negligible.

In Fig.~\ref{fig:2dF}, we show the distribution of star-forming
properties of isolated galaxies and isolated pairs measured via this
technique from the 2dF survey \citep{colless01}.  The close ($\leq
50$~h$^{-1}$ kpc) pairs show a distinct, 14\% excess of galaxies with
high star formation rates.  In \citet{barton07}, we show that the
ability to measure this effect accurately depends dramatically on whether pairs
in denser environments are removed from the sample.

\section{Widely Separated Pairs and the Evidence for Quenching and ``Strangulation''}

\begin{figure}
\plottwo{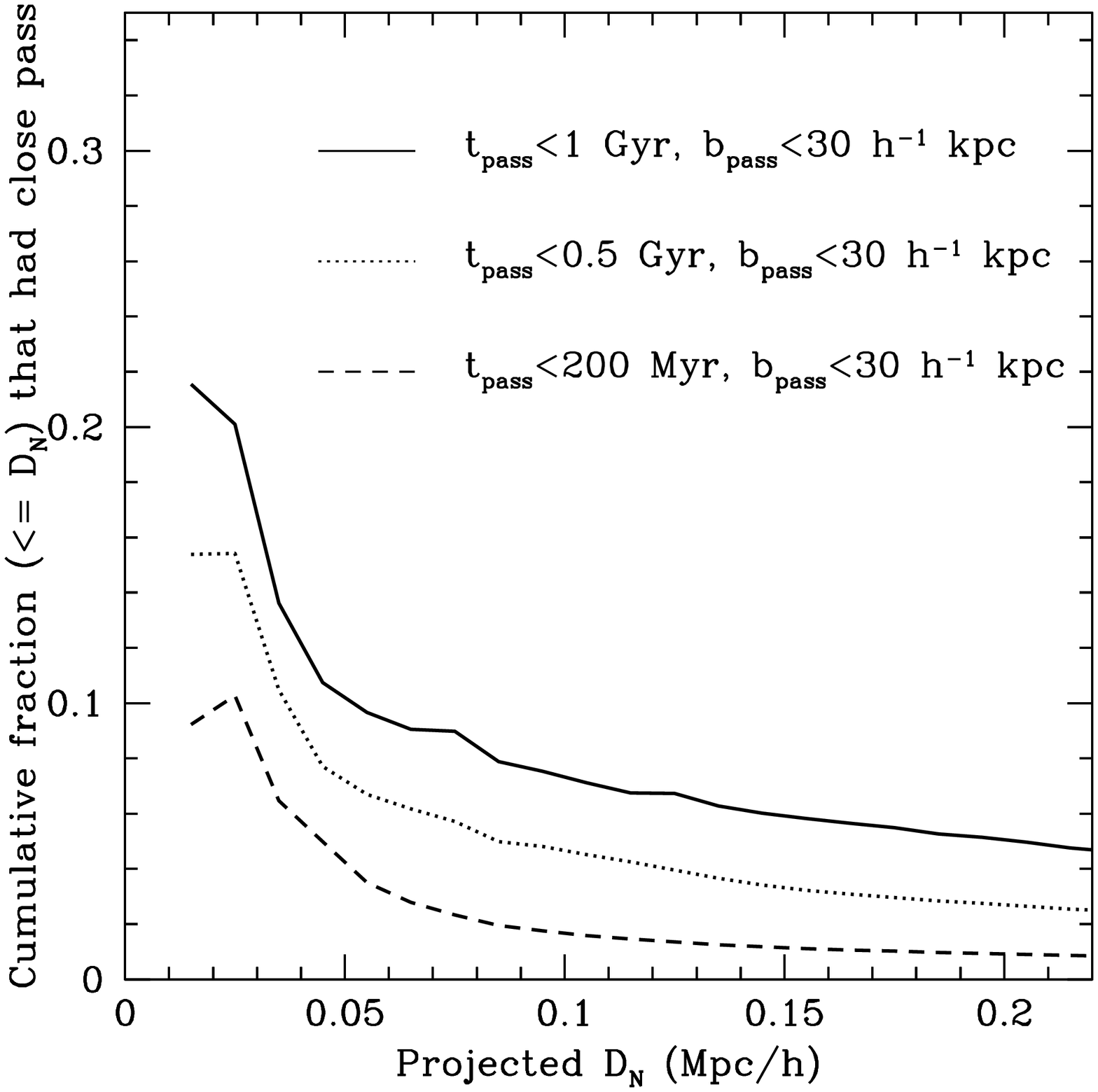}{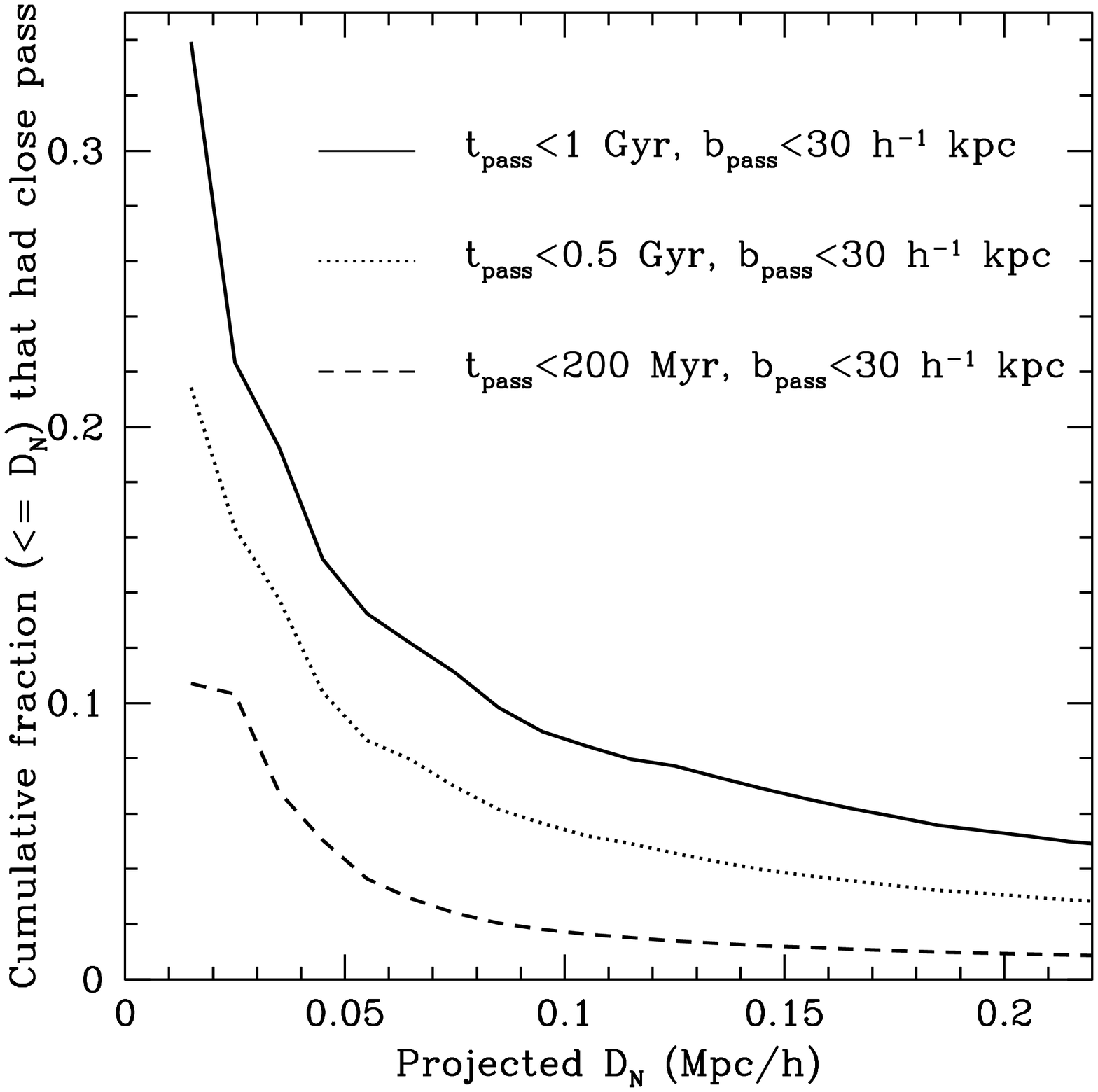}
\caption{The cumulative fraction of isolated pairs that have had a
close ($\leq 30$~h$^{-1}$ kpc) pass recently, where ``recently'' is
either 1 Gyr ({\it Solid line}), 500 Myr ({\it Dotted line}), or 200
Myr ({\it Dashed line}), as a function of the distance to the nearest
neighbor.  We plot the fraction for two different cutoffs, including
({\it Left}) all halos with V$_{\rm max} \geq 160.5$~h$^{-1}$ kpc,
appropriate for comparison to the 2dF work in \citet{barton07}, and
({\it Right}) with V$_{\rm max}$ $\geq 140$~h$^{-1}$ kpc, appropriate
for comparison to a volume limited SDSS sample with \Mrh\ $\leq -19$
\citep[e.g.,][]{trinh09}.  The fraction of pairs that have had close
passes recently is a strong function of \DN.}
\label{fig:passes}
\end{figure}

While close pairs are ideal for the study of triggered star
formation, more widely separated isolated pairs reveal
other group processes.  Most galaxies in systems with \Nhalo\ $=2$
members have actually not had a recent close galaxy-galaxy pass.  In
Fig.~\ref{fig:passes}, we use the simulation to compute the cumulative
number of galaxies that have had a close pass ``recently'' (200 Myr to
1 Gyr ago), for two different halo masses (or V$_{\rm max}$ values),
corresponding to two different magnitude limits for the data.  In both
cases, the fraction of galaxies that have had a close pass is a strong
function of the distance to their nearest neighbor, becoming $\leq
10$\% for pair samples with maximum separations above \DN\ $\sim 0.1$
h$^{-1}$ Mpc.  As a result, more widely separated pair samples can be
used to study whether the processes that cause the morphology-density
relation \citep[e.g.,][]{dressler80,postman84} operate in extremely
sparse systems that have only two members.  These processes include
rapid quenching, where the cold gas is stripped directly from the
system, and ``strangulation,'' where only the rarefied outer halo is
removed.

\begin{figure}
\plottwo{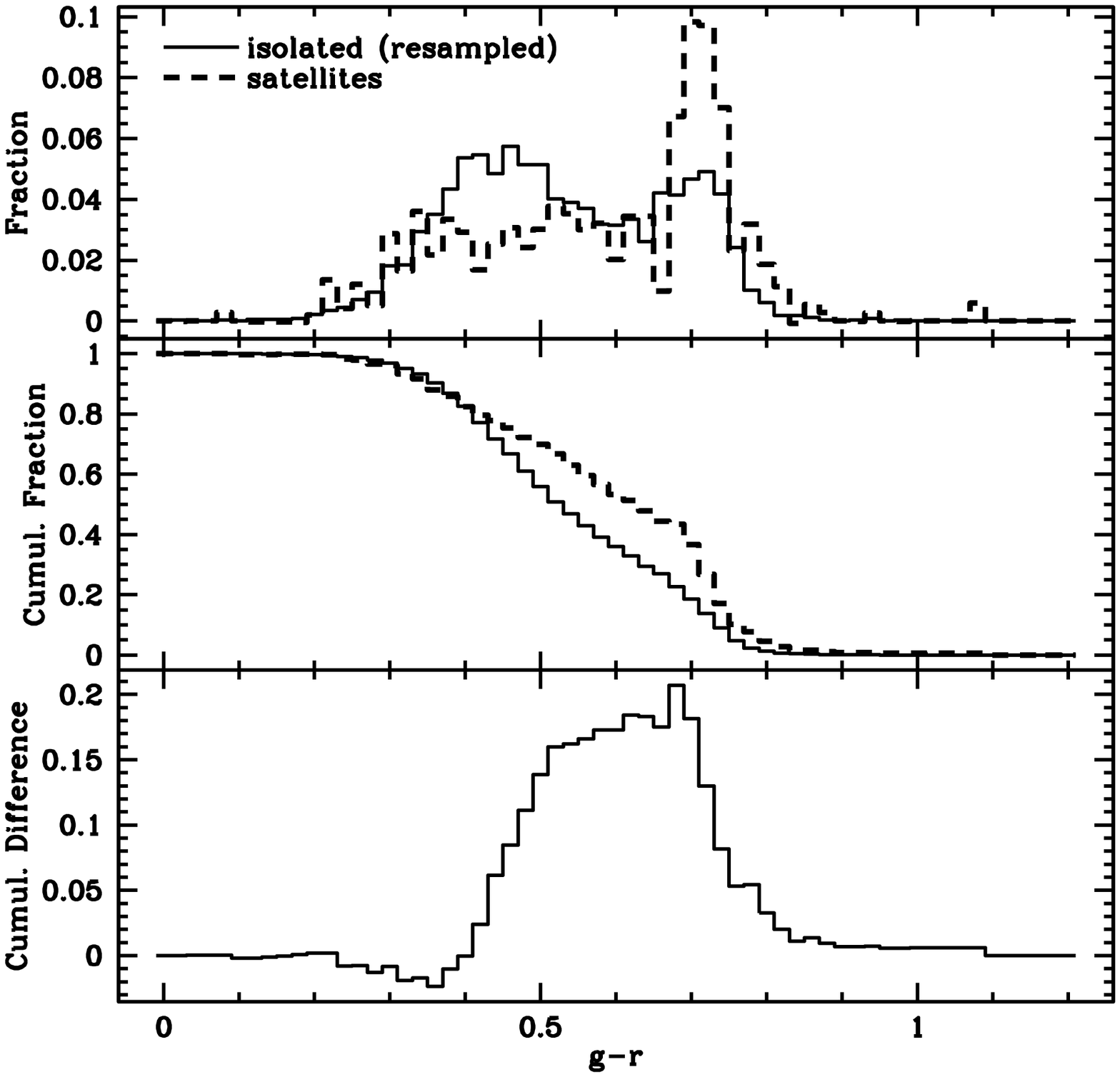}{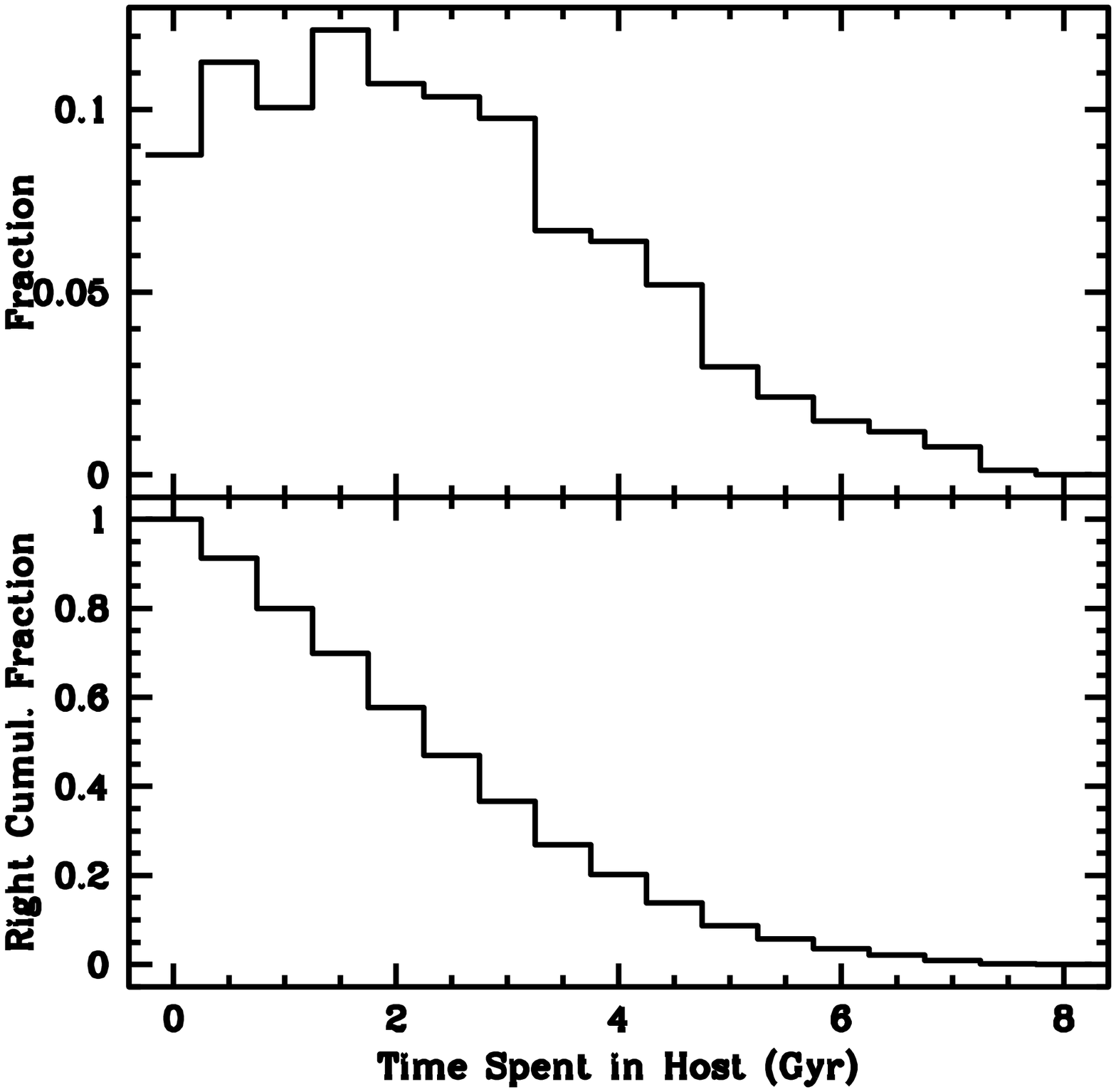}
\caption{Widely separated pairs from the SDSS, adapted from 
\citet{trinh09}.  ({\it Left}) We plot
histograms of $g-r$ color for isolated galaxies and satellite galaxies
in pairs with \DN\ $\leq 200$~h$^{-1}$ kpc ({\it Top}), along with
the cumulative distributions ({\it Middle}) and the difference between
the cumulative distributions ({\it Bottom}).  The pairs have had
contamination from isolated galaxies subtracted and the isolated
galaxies have been resampled to the same stellar mass distribution as
the satellites.  ({\it Right}) From the simulations, we plot the
differential ({\it Top}) and cumulative ({\it Bottom}) distributions 
of the amount of time that subhalos of \Nhalo\ $=2$
systems have spent as substructure.}
\label{fig:chris}
\end{figure}

In Fig.~\ref{fig:chris}, we demonstrate the difference in color
distributions between the wider pairs and isolated galaxies of the
same stellar mass.  The satellite galaxy is defined as the less
luminous member of each pair ($r$-band).  The satellites show a slight blue
excess, presumably due to triggered star formation, and a much
stronger red excess due to quenching and/or ``strangulation.''
Specifically, 21\% more satellites than isolated galaxies are redder
than $g-r=0.68$, {\it even when galaxies with the same stellar masses
are compared.}  Thus, 21\% of satellite galaxies in even the tiny
\Nhalo\ $= 2$ systems have moved from the blue to the red sequence.
At least some of the processes that give rise to the
morphology-density relation operate in even very sparse groups.

The right side of Fig.~\ref{fig:chris} uses the simulation to measure
the amount of time that the satellites (subhalos) have been
substructure in these \Nhalo\ $=2$ systems.  Almost independent of star
formation history, it takes roughly 2 Gyr to go from blue to red when
star formation stops. If star formation ceased as soon as the galaxy
became substructure, roughly $\sim$60\% of all satellites would have
stopped forming stars.  Because $\sim$77\% were blue to begin with,
roughly $\sim$46\% of the satellites would have moved from blue to
red.  Because the number that have actually moved from blue to red is
only 21\%, we conclude that star formation continues well after a
galaxy becomes substructure, for typically $\sim$2 Gyr.  This analysis
yields significant constraints on quenching and ram pressure stripping
outside of the cluster environment \citep{trinh09}.

\begin{figure}[t]
\plotone{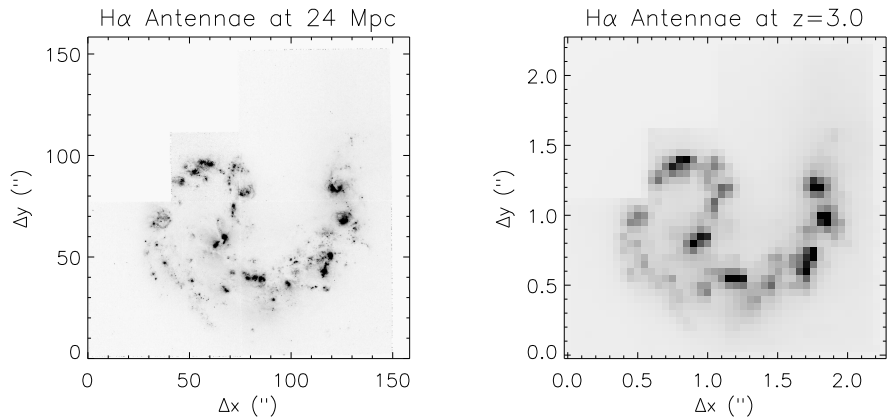}
\caption{A simulation of the Antennae observed at $z=3$.  We show an
H$\alpha$ image of the Antennae courtesy of B. Whitmore ({\it Left})
and a simulation artificially redshifted to $z \sim 3$ and observed
with the an integral field unit spectrograph on the Thirty Meter
Telescope ({\it Right}) with an exposure time of 3 hours and a
sampling of 50 mas.  Individual star-forming lumps are
discernible with TMT and will yield an accurate kinematic map of the
system.}
\label{fig:TMT}
\end{figure}

\section{Detailed Studies at High Redshift: the Next Generation of Large Ground-based
Telescopes}

With present-day technology, the signatures of interaction are
difficult to identify unambiguously at high redshift.  New kinematic
techniques enabled by adaptive optics and integral field unit
spectrographs are effective
\citep[e.g.,][]{wright09,forster09}. However, these observations are
limited to the brightest systems and do not reveal individual
star-forming clusters.  In the era of the next generation of large
optical and infrared telescopes, adaptive optics and huge mirror sizes
will allow a very detailed identification of merging systems.  In
Fig.~\ref{fig:TMT}, we present a simulation of the Antennae galaxies
observed with the Thirty Meter Telescope (TMT) and the InfraRed Imaging
Spectrograph (IRIS), which is planned to work with adaptive optics at
first light. The observation reveals individual superstar clusters
even at $z \sim 3$, and likely beyond, allowing very detailed kinematic and metallicity maps of the
galaxy.

\acknowledgements 

We would like to thank our collaborators, Risa H. Wechsler and
Andrew R. Zentner, for allowing us to quote these results prior to
publication. We would also like to thank Brad Whitmore for sharing his
reduced H$\alpha$ image of the Antennae.


\end{document}